\documentclass[conference]{IEEEtran}
\IEEEoverridecommandlockouts
\usepackage{cite}
\usepackage{amsmath,amssymb,amsfonts}
\usepackage{algorithmic}
\usepackage{graphicx}
\usepackage{textcomp}
\usepackage{xcolor}
\usepackage{listings}
\usepackage{subcaption}
\usepackage[font=small,labelfont=bf]{caption}

\usepackage{setspace}
\IEEEaftertitletext{\vspace{-2\baselineskip}}

\def\BibTeX{{\rm B\kern-.05em{\sc i\kern-.025em b}\kern-.08em
    T\kern-.1667em\lower.7ex\hbox{E}\kern-.125emX}}
\begin{document}

\title{Chiplets and The Codelet Model\\
{}
\thanks{This research was supported by the Exascale Computing Project (17-SC-20-SC), a collaborative effort of the U.S. Department 
of Energy Office of Science and the National Nuclear Security Administration. We gratefully acknowledge the computing resources provided and operated by the Joint 
Laboratory for System Evaluation (JLSE) at Argonne National Laboratory. This work is partially supported by the National Science Foundation, under award SHF-1763654. This is a part of Project 38, a joint DOE-DOD effort to identify new breakout architecture features for memory systems}
}

\author{\IEEEauthorblockN{Dawson Fox}
\IEEEauthorblockA{ \textit{Argonne National Laboratory, Lemont IL}\\
                    \textit{University of Delaware, Newark DE}\\
                    \textit{dawsfox@udel.edu}}
\and
\IEEEauthorblockN{Jose M Monsalve Diaz}
\IEEEauthorblockA{\textit{Argonne National Laboratory} \\
\textit{Lemont, IL}\\
\textit{jmonsalvediaz@anl.gov}}
\and
\IEEEauthorblockN{Xiaoming Li}
\IEEEauthorblockA{\textit{University of Delaware} \\
\textit{Newark, DE}\\
\textit{xli@udel.edu}}
}

\maketitle

\begin{abstract}
The past decade has seen a rapid evolution in hardware technology computation. New architectures that promise to improve performance for domain-specific applications have started to appear. In the future, processors may be composed of a large collection of vendor-independent IP that specializes hardware for application-specific algorithms, resulting in extreme heterogeneity. However, integration of multiple vendors within the same die is difficult. Chiplet, a promising technology, provides a packaging method that integrates multiple vendor dies within the same chip by breaking each piece into an independent block. A common interconnect allows fast data transfer across chiplets. 

Most research on chiplets has focused on the interconnect technology. In comparison, programming the chiplet architecture is a less-touched question. Our vision is that it is necessary to define a program execution model (PXM) that enables programmability and performance in chiplet architectures, especially in the  handling of multiple vendor-independent chiplets. The key is that a cohesive co-designed PXM can effectively separate the roles of the different actors, while maintaining a common abstraction for program execution.  This position paper describes how co-designed Program Execution Models can be used in chiplet architectures. Additionally, it proposes the Codelet PXM as a feasible candidate for a general purpose program execution model in extremely heterogeneous chiplet-based architectures. This paper also presents preliminary examples of how using the Codelet Model can help performance through architectural features and performance models.
\end{abstract}


\section{Introduction}
For a long time, the computing industry relied on advancements
within sequential processors and hardware for better performance. This approach first hit the
power wall, which led to the adoption of multi-core shared memory architectures. That architecture is
essentially a collection of Von Neumann machines linked by a unified view of memory, each capable of executing 
independent instructions in parallel and using concepts such as mutexes to synchronize. 
Unfortunately, the force behind this transition resulted in both extra
hardware to resolve issues that arise within parallelism (e.g. enforcing memory consistency models, cache
coherency through invalidation mechanisms, atomic memory operations and their related overheads) as 
well as increased development effort and less predictability on the software side. Now, as Moore's law
slows down, the industry 
has turned towards heterogeneous architectures that better suit the mix of workloads. This can be seen with
the growing prevalence of accelerators such as GPUs, TPUs \cite{TPU}, ML/NN/AI accelerators \cite{UsingSambaNova}, or specialized sparse
computation accelerators. An example of accelerators' presence in the industry is Apple's use of the M1 and M2 chips \cite{appleM1}\cite{appleM2},
which contain RISC cores, GPU cores, and neural network cores all on the same silicon. Furthermore,
cost and power efficiency continue to be paramount in the HPC field due to the necessarily larger
scale of the systems involved. Cost and the need for a variety of application-specific architectures
creates the need for customized chips; after all, HPC systems have specific needs that cannot all be fulfilled 
by one vendor.

A solution that
has been proposed and appears to hold promise for the near future is the idea of chiplets: each
chip containing smaller portions of architectures specialized for a single purpose or field. Chiplets provide
a cost-effective way to manufacture smaller, specialized portions of silicon that can be grouped
into a single chip, providing the flexibility of integrating different vendors and the heterogeneity to fulfill the specific needs 
a system might have. 
However, while the clear benefit of specialized architectures has been proven in a variety of cases, the
industry is sorely lacking a cohesive way to organize the execution of codes between highly
differing hardware within the same silicon, as well as methods to optimize memory access and data movement
within such a system. Recent research in this area has been heavily focused on network-on-chip (NoC) / interconnect and routing algorithms \cite{ModularRoutingDesign} between chiplets or
specific chiplet layouts for a particular implementation \cite{AMDZen2} \cite{SimbaChiplets}, \cite{96Core6Chiplets}.
While the format of chiplets has previously been explicitly mapped to specific computational problems
with due consideration given to software complexity such as in \cite{Chien10x10}, there is a significant
gap left between hardware/architecture/interconnect design and software toolchains that are necessary for development.
Given that chiplets in a custom chip will likely be developed by different vendors, this gap will widen if a sound program execution
model is not selected as a general vision of the system's architecture. 

Meanwhile, 
parallel architectures are generally left with conventional memory hierarchies. 
Caches began as an excellent invisible optimization for general sequential program execution and were
later adapted to provide correctness and programmability in parallel systems. The sequential past of these designs force parallel programmers to contend with
ambiguous memory location and high cache invalidation overheads. Today, taking into consideration
the memory wall, some applications that particularly suffer from the built-in protocols of the
cache, cache coherency, and prefetching mechanisms use unique memory hierarchies in their corresponding 
specialized architectures \cite{Sparch}\cite{OuterSpace}\cite{MetaStrider}. Techniques such as more intelligent prefetching, smarter memory management
based on the execution model, and data streaming through FIFOs can ease this burden more generally.


At a larger scale, the need for load balancing and handling massive amounts 
of data further drives the need for well-organized
execution and memory movement through and between cache hierarchies. 
This need leads us to the concept of a program execution model (PXM), which already exists
in the use of any parallel execution, but may not be explicitly represented. PXMs currently in use do not have clear definitions nor hardware support for their operations, causing them to ultimately fall short of their potential. 
We view the PXM as the key concept that chiplet architectures need to provide the missing flexibility and programmability. In particular, we propose the use of the Codelet Model PXM, a well-defined fine-grain event-driven PXM,
for computation on heterogeneous chiplet-based systems.



This paper aims to convince the reader that, in order to achieve programmability, performance and portability, it is necessary to define a sound program execution model that acts as a contract between users, developers, software infrastructure and hardware designers. In order to achieve this, this manuscript answers the following questions:
\begin{itemize}
    \item Section \ref{motivating example} answers the question \textbf{What problems are likely to be seen when programming on systems using chiplets from various venders?} This section presents a brief example of difficulties that users and developers may experience.
    \item Section \ref{pxm}s answer the questions \textbf{What is a Program Execution Model?} and \textbf{Why do we need to define one?} This section describes PXMs and their potential effects on performance, efficiency, and programmability.
    \item Section \ref{codelet model} answers the question \textbf{What is the Codelet PXM?} This section describes the Codelet Model, its prior implementations, and its effect as a fine-grain event-driven PXM. 
    \item Section \ref{codelet model for chiplets} answers the question \textbf{Why is the Codelet PXM a suitable model for Chiplets?} expressing driving reasons to use the Codelet Model as a PXM on chiplet based systems.
    \item Finally, Section \ref{case study} uses a simple Matrix Multiplication case study to demonstrate how the Codelet Model abstraction enables the expression of problems and the mapping of implementation to an abstract machine that may evolve over time.
\end{itemize}
\section{Motivating Example} \label{motivating example}

The problem of not having a well-defined program execution model is inspired by software, but affects all the layers of computer systems. Software has the ability to evolve faster than hardware, and it also allows fast creation and modification of machine abstractions and their behavior. Consequently, current PXMs for parallel computing have been mostly implemented in software, with little direct support in the hardware design.


For programming homogeneous parallel machines (i.e. Multi-core CPUs), the thread abstraction was a natural evolution of software for a von Neumann based multi-core hardware. Previously, it has been discussed in \cite{LeeProblemWithThreads} how threads are difficult to program and highly non-deterministic. As a result, other parallel abstractions have been built using threads as building blocks. Each newly introduced execution model imposes a machine abstraction and a view of workers and work assignment. For example, OpenMP exposes not only a fork-join model but it also allows users to develop programs in the form of tasks. The former sees the system as a collection of threads with fixed work assignment, while the latter includes queues and schedulers and separates threads from the unit of computation, i.e. the task. Other more complex programming frameworks such as Legion\cite{BauerEtAlLegionExpressingLocality2012}, Raja\cite{Raja}, Kokkos\cite{edwards2014kokkos}, DARTS\cite{DARTSEuropar}, and others, have different machine abstractions as well, some more complex than others.

With the arrival of heterogeneity the execution models were extended to account for the new resources and their coordination. Currently, the most commonly used model for accelerators is an offloading model, where the host orchestrates execution of tasks (i.e. kernels) in the accelerator device. Each device may feature a different execution model, and additional runtimes may be needed to bridge the gap between them. We do not propose the removal of software runtimes, but we argue that not having a common PXM leads to potential conflicts across vendors that lead to performance degradation. Furthermore, not solving this missing feature reduces the chance to have hardware innovation of PXMs implementations that lowers runtime overheads.

Let us imagine that we have a system that contains 5 chiplets, each with a different architecture. Currently, with an offloading model that is exposed as an API to the user, we will see the code similar to the following:

\begin{lstlisting}
    host_code() {
      void * A, B, C, D, E;
      B = Chiplet1_API(A);
      C = Chiplet2_API(B);
      D = Chiplet3_API(B);
      E = Chiplet4_API(C,D);
      Chiplet5_API(E);
    }
\end{lstlisting}

We are using C-like syntax as a commonly-used low level interface in these devices. And the \texttt{void *} to represent any data type that operates on these chiplets. Inside each API call, there exists a runtime that prepares the device kernel and submits jobs to it. Consider also that these chiplets are from different vendors, meaning that each vendor is in charge of developing its runtime system. Following, we present some problems that arise from the lack of a common PXM.

\paragraph{\textbf{Multiple runtime systems}}
Imagine that each company decides to use a different parallel programming model. For example, Chiplet1 vendor uses OpenMP, Chiplet2 vendor uses OpenACC, Chiplet3 vendor uses a custom made model, and so on. Problems arise from possible collisions between these runtime systems. Each runtime system assumes that they are sole owners of the system, resulting in conflicts in resource utilization. At best, these conflicts only incur performance penalties. However, since different runtimes do not necessarily share their state information, there exists a chance that conflicts lead to erroneous behavior in the program execution. Furthermore,  across vendors, there's not enough information on how these APIs are any different to any other function call, resulting in lack of interoperability. In general, unless a compiler is thought to understand all the possible APIs, it is not possible to obtain semantics of the execution model through API calls.
\paragraph{\textbf{Unified shared memory abstractions}}
Many runtime systems and chiplet-based architectures are vouching for unified shared memory abstractions, arguing that it makes the code more portable and the system more programmable. However, unified shared memory tampers with the possibility of dataflow analysis and pointer-based optimizations across compilation units. In the example, there's no guarantee that any of the pointers (A through E) have no aliasing, or that there are hidden data dependencies between multiple API calls (e.g. variables in global memory). Isolation of runtime development across vendors under a unified shared memory, without a cohesive view of the system (i.e. PXM), exacerbates this problem. On the other hand, a pure von Neumann view of the system requires that memory and state are synchronized across different devices. Coherency becomes a fundamental element to support this mechanism but is expensive and difficult to scale. Furthermore, a pure cache-based system misses on performance opportunities that arise from manual memory management. To solve this issue, it is possible to add user managed memory (e.g. scratchpad memory). But it burdens the user once more with memory management, removing some of the programmability gained from a unified shared memory system. While unified shared memory may be a good solution to portability of old code, its adoption is costly and restrictive of more relaxed memory models presented by a PXM. 
\paragraph{\textbf{Data, dependencies and parallelism}}
Firstly, data may require different representations on each chiplet, increasing the overhead of data transformation from one format to the other. Possible additional calls or logic may be needed between each API call in our example to adapt data to the right format (e.g. from column major to row major). Furthermore, in order to properly take advantage of chiplets, parallelism is necessary. The user, or another programming model, is now required to create a new execution model on top of the chiplets API that allows for properly managing execution. Nevertheless, a software only approach is yet another runtime system that needs to be supported to connect all the chiplet's runtime system. An explicitly defined PXM that serves this role will lead to hardware improvements and reduce the cost of interconnecting different chiplet's execution models. 

To this end, tasking, is a proper abstraction that allows for management of dependencies and parallelism. However, there is currently no tasking PXM that is widely acceptable as a candidate, and, as previously mentioned, a pure implementation in software at the user level will likely generate more collisions with the runtimes used in each chiplet vendor. A generalized PXM can lead to improvements in hardware and Operating System runtimes that free vendors and users from designing these models.
\paragraph{\textbf{Missed opportunities}}
Finally, optimizations across vendors are often missed due to the lack of agreement between different runtime systems and abstractions. For example, allowing streaming and pipelining across vendors \cite{Popovici2021ImprovingDataLocality} is difficult to achieve in the example code above. It is possible to build software abstractions on top of these API calls, but again it is left to the user or runtime system used by the user. Pointers do not properly expose operation sizes, which does not allow for load balancing. APIs have little information about memory access order, leaving users and compilers without opportunities to improve performance. By using a pure API abstraction, with little information on the execution model used across the program, we are missing on opportunities. 

We will now take a deeper look at PXMs and their role in programming and execution.
\section{Program Execution Models} \label{pxm}

A PXM has been originally described as the “formal specification of the application program interface (API) of the computer system” \cite{Dennis1997}. This brief definition conveys that the PXM is relevant
throughout the entire system and that it is necessary to operate it correctly. Dennis et.al. \cite{Dennis1997} also states that a PXM sufficiently describes the system’s semantics for compiler writers as well as
providing a goal for designers of a system itself. The PXM in this way constitutes an agreement between hardware and software much like an ISA, but where an ISA is concerned with a single computational unit (a
single core, for example), the PXM is system wide. PXMs can be broken down into three models according to \cite{ParallelTuringMachine}: an activity model, a memory model, and a synchronization model. The activity
model defines the permitted parallel activities that can occur on the system (in many systems for example, this may be the threading mechanism). The memory model describes how memory is addressed as well as the
effects and results caused by operations on the memory. The synchronization model describes the behavior when parallel activities and the memory interact, including the memory consistency model of the system (which
on conventional multiprocessors is often realized as Lamport’s sequential consistency \cite{SequentialConsistency}). While all parallel execution invokes a PXM of some sort, modern PXMs currently in use are
limited.

 \begin{figure}
     \centering
     \includegraphics[width=0.4\textwidth]{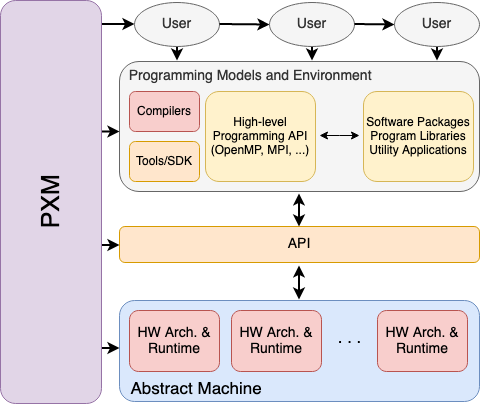}
     \caption{A Program Execution Model implemented in a heterogeneous system. Note that the PXM entails both the programming model and the hardware, with an API to utilize hardware in the form of the corresponding abstract machine. Particularly, although OpenMP implements its own PXM, the coverage is limited to the high-level programming level of the system.}
     \label{fig:hetero_pxm}
 \end{figure}
 
Much of modern PXMs’ limitations are caused by lack of flexibility or their implementations. PXMs are not necessarily confined individually to hardware or software (see fig. \ref{fig:hetero_pxm}). It can be thought of as the
API to a specified abstract machine, and therefore the implementations can vary widely between the hardware directly supporting the abstract machine’s model of execution or being implemented solely in
software to fit hardware that does not match the abstract machine at all. A modern day example of this is OpenMP, which implements the OpenMP PXM in software. This can be seen in that not all essential
synchronization operations/interactions are supported in hardware. Thus, thread creation, tasks, barriers, and synchronizations are not hardware elements but rather software constructs. However, these operations are crucial for correct execution on the OpenMP PXM and yet exist only in software.

On the other hand, when a PXM is implemented or supported in hardware, it still contains components from various levels of the computer system because it represents the overall behavior of the system
as a whole when performing parallel computation. True to its definition as a specification of the API of the computer system, it provides procedures that users can invoke, which naturally occurs at
the software level. Continuing with OpenMP as the example, one can see how OpenMP clearly provides an API for users to execute parallel programs on a system with correct behavior. The central problem
for this section is that, much like OpenMP, modern PXMs are forced to be implemented in software on top of von Neumann abstractions, which inherently limits their performance. 

The conventional view of computer systems is too fragmented to facilitate a clear model of parallel computing. The overhead of a PXM is much greater and sometimes prohibitive when implemented in such
a fragmented manner. Modern PXMs are built to provide efficient and effective parallelism for user code but forced to rely on interfacing behind the scenes with multiple computational elements that
are by and large individual von Neumann machines with a shared memory abstraction. Furthermore, the popular model of parallelism wherein multiple threads have unrestricted behavior and interact in
mysterious ways leads to confusion and can make programs near-impossible to debug \cite{LeeProblemWithThreads}. 
Furthermore, data is moved throughout the memory hierarchy and between computational units, largely hampering non-conventional programs and giving little control to user code.
The goal of PXMs is the pan-system agreement between all software and all hardware of the system, though they currently fall short of this goal.
This agreement should be strict enough to build reliable and understandable programs upon, but flexible enough to allow ease of programming and composability. It should also remain implementation
independent (hence, vendor independent), so that innovation in hardware and software do not necessarily have to be dependent on each other. The PXM focuses on interaction of components, so as long as the semantics
are enforced, the computational units themselves can be self-contained and need not have any specific execution model within them (note the variety of hardware architectures and runtimes in fig. \ref{fig:hetero_pxm}).

In short, a well defined PXM will remove the decidedly unpredictable behavior that so commonly appears in developing and executing parallel applications, and a holistic PXM implementation will
support better performance and efficiency. While PXMs currently are lacking as described, the issues will only grow with the arrival of chiplets. PXMs that already result in unpredictable programs
will only become worse, as mentioned in section \ref{motivating example}. When it comes to utilizing multiple architectures with a single program (heterogeneity),
the PXMs in use are mostly offloading-based. This model is inherently limiting with regards to achieving
high utilization amongst a variety of architectures, most likely involves explicit synchronization on the developer's part, 
and can include kernels with unrestricted behavior (possible state change, accesses to
global memory, etc.). In a chiplet-based system where these varying
architectures are necessarily smaller so as to fit multiple on a single chip, it will be even more difficult to efficiently utilize the components in conjunction with one another.
Ideally, a PXM for a chiplet system would include clear definition of behavior, input, and output of workloads. Beyond this, it would provide clear interactions between parallel
workloads and memory to achieve more understandable and predictable execution. To this end we propose the Codelet Model PXM, opening the door to better software analysis/optimization,
improved isolation, memory efficiency and utilization.

\section{Codelet Model}  \label{codelet model}
\begin{figure}
    \centering
    \includegraphics[width=0.4\textwidth]{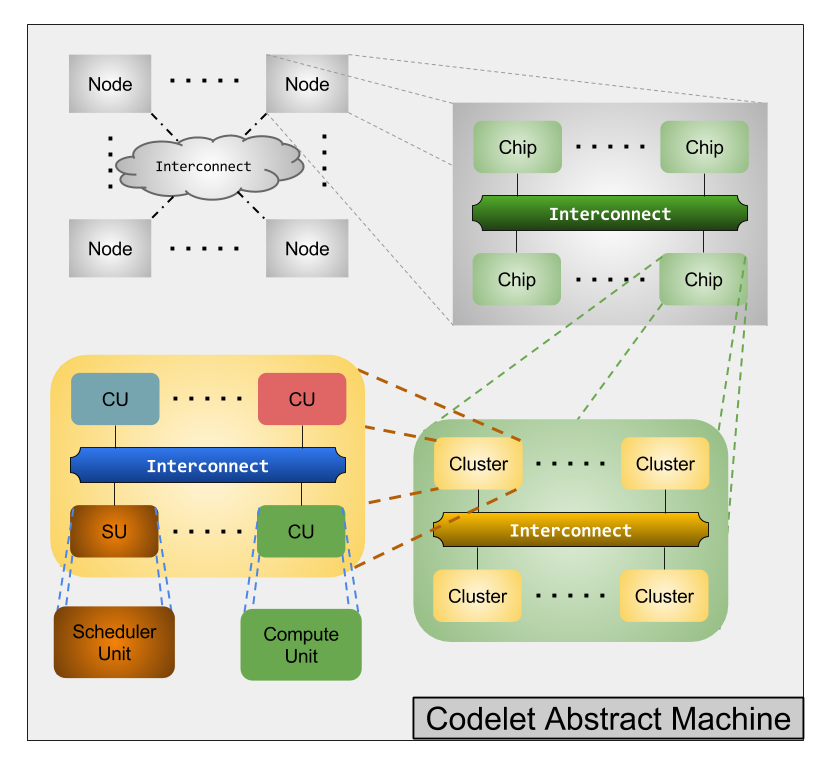}
    \caption{A hierarchical, generic representation of the basic Codelet Abstract Machine. A Cluster contains an SU and an arbitrary number of CUs. TPs belong to a specific cluster for locality. \cite{ParallellaCodeletModel}}
    \label{fig:cam}
\end{figure}

The Codelet Model is a dataflow-inspired PXM to organize computation with an accompanying abstract machine. It is both fine-grain and event driven, breaking computation into Codelets, portions of 
sequential computation with defined inputs and outputs that are non-preemptive. As such, Codelets are the quantum unit of scheduling of a Codelet-based program. Programs in the 
Codelet Model are described by a Direct Acyclic Graph, with nodes in the graph representing Codelets
and directed arcs representing data dependencies between them. This allows the program to clearly define the
ordering of Codelet execution only where necessary while permitting flexibility in Codelet scheduling
otherwise. To benefit more from locality, Codelets are grouped into Threaded Procedures (TPs). The
Codelet Abstract Machine defines the components of a system that executes Codelet Programs
(see fig. \ref{fig:cam}).
It then designates the roles of Compute Unit (CU) and Scheduling Unit (SU):
the SU is responsible for creation of TPs and scheduling Codelets during runtime, while the CU
is responsible for "firing" (executing) Codelets once their dependencies are fulfilled. SUs and CUs
are grouped into "clusters"; the cluster designation is important because TPs can only belong to a 
single cluster, thus providing the locality mentioned. Codelets, when they have unfulfilled
dependencies, are considered dormant. Once their dependencies are all met, they become enabled. When
a CU is available to execute the Codelet, the Codelet is considered ready and, having begun execution,
is called active. After completing execution, the Codelet can be reset and return to the dormant state,
easily facilitating reuse.

For those new to the Codelet Model, a common question is "how are Codelets different from threads?" A
very simple answer is that Codelets are lightweight, non-preemptive and stateless. A user may write a thread
that yields and later resumes execution for example; by definition, a Codelet may not cease execution
until it is complete, thus constituting the principal scheduling quantum (PSU) of the Codelet Model
\cite{ParallelTuringMachine}. In accordance with this principle, a Codelet also cannot be migrated from one CU to another
while executing. Because Codelets are stateless, they act as general procedures that can be fired
multiple times with varying input and should never incur the overhead of saving state. Codelets do not need to save state since they
always execute until completion on a single CU and leave behind no state to be saved.

Further questions may include, "how is a Codelet different from a task?" The main difference here
is that Codelets have well-defined input and output and are side-effect free. This means that under
the Codelet Model PXM, the entire system can rest assured that only the data marked as accessible
by the definition of the Codelet will be accessed by it. On the other hand, tasks are often described
as a function pointer, a private data environment and its dependencies, with unrestricted function size
and unfettered access to global state. Furthermore, because the Codelet Model was designed as a PXM and
the Codelet has a clear definition in terms of behavior, a system could be designed where the concept of
a Codelet is understandable even from a hardware perspective which opens the door to low level 
optimizations and hardware supported mechanisms. These characteristics of the Codelet Model form its core
as a PXM, but have been extended to expose more benefits in the past.


\subsection{Codelet Model Extensions}
A very straightforward extension to the Codelet Model is adding conditions that must be met for a
Codelet to be enabled. For example, in prior
work, "just-in-time locality", i.e. the necessary data being local to the unit
the code is executing on just before execution begins, was a necessity \cite{GarciaOroKhaVenLivGao12} that could be
implemented in the Codelet Model as a new type of dependency. 
This concept helps mitigate the loss of potential latency hiding that is caused by 
Codelets being non-preemptive. In fact, the well-defined inputs and outputs of each Codelet present
the opportunity for smarter data movement, though this is implementation dependent and not
specific to the Codelet Model's PXM. In another case, it is possible that a Codelet requires access
to a shared resource; it could feasibly have as a dependency that the resource is available and could
rely on this dependency being fulfilled before firing similar to a typical data dependency.

In addition to extensions of the meaning of a Codelet being enabled (as above), the Codelet Model
could be extended to benefit from heterogeneity in parallel \cite{TongshengPDAWL2020}.
Traditionally, while a Codelet Model "cluster" may not contain only identical CUs, CUs have been viewed
as identical from the perspective of the SU. However, as long as the system throughout supports the
mechanisms of signaling between Codelets to track whether or not dependencies are met, and as long
as Codelets are written in a way such that they can be correctly executed on a given CU, the Codelet
Model will have correct execution. A straightforward example of this would be if multiple CUs have
the same ISA but differing memory hierarchies (different sized caches, cache lines, etc.). Writing the
Codelets would not necessarily require any changes; however, with added metadata to indicate targeting
a certain CU and with either compiler optimizations or an experienced developer, more efficient
execution could be achieved. Now we will explore some of the benefits that have been gained from using
the Codelet Model in the past.

The Codelet Model has also been used to define a hierarchically structured von Neumann machine\cite{MonsalveIDPRM2019}. 
A program for this machine abstraction is envisioned as a hierarchical description of Codelets. 
Parallelism is achieved through techniques inspired by instruction level
parallelism such as out of order execution engines and superscalar architectures \cite{MonsalvePhDThesis}. 
This work also considers heterogeneity, and enables compiler optimizations to take advantage of program execution \cite{monsalveExHET}.

\subsection{Prior Codelet Model Implementations}
DARTS (Delaware Adaptive RunTime System) is the basic software-level Codelet Model implementation
that is the basis of prior work on improving parallelism and efficiency using the Codelet Model
\cite{DARTSEuropar}. More specifically,
various Codelet Model implementations have already been shown to offer benefit to conventional 
computations. Take, for example, the use of the Codelet Model to compute fine-grain FFTs \cite{YaoWuThesis}.
The advantage of the Codelet Model's ability to provide fine-grain parallelism can help achieve better
utilization and load balancing. In addition, \cite{YaoWuThesis} is a pertinent example of how
the clear memory inputs and outputs of Codelets can be leveraged to optimize memory accesses,
balance the application, and further exploit locality. A fine-grain Codelet Model implementation was
also used to target parallel stencil computation in \cite{GengEtAl16}. It can
be seen there that speedup was achieved over coarser-grain models, but the limitations of a software
only PXM can be noted as well: the performance was somewhat dependent on von Neumann cache behavior and platform
dependent mechanisms. This once again indicates the need for a cohesive PXM throughout the varying
levels of the system to make advancements in performance and efficiency generally which could only
otherwise come from platform dependent implementations. An example of an extended Codelet Model
implementation can help us demonstrate this.

DARTS has previously been used to include streaming as a method of 
performing dataflow software pipelining at the Codelet level \cite{SidThesis}. The implementation
used FIFO queues to perform the pipelining and was tested on GEMM using Cannon's algorithm; it was
shown to provide speedup, better computational efficiency, and lower synchronization overhead than
the non-pipelining implementation. There is undergoing work to integrate the implementation of 
\cite{SidThesis} into the runtime itself as an effort to increase programmability and generality
of the streaming mechanism and apply the mechanism to more complex computational problems. 
The streaming Codelets could then be assigned the FIFOs by the scheduling
mechanism and access them only using typical push and pop methods, effectively providing fine-grain
pipelining between them at very low effort to the application developer. This 
demonstrates well the potential the Codelet Model PXM misses out on without hardware support: on a typical
von Neumann multicore machine, without a HW implemented FIFO, it is completely ambiguous to the 
runtime and to the application developer where the FIFO is physically stored, and it may be affected
unpredictably by mechanisms like cache coherency and false sharing.

Unlike the addition of dataflow software pipelining which mostly provides better performance at
the cluster level, advancements can also be made by extending the Codelet Model to handle system wide
scheduling better. Within the Codelet model, there is a native two level hierarchy of computational units: the Codelet,
which is the principal scheduling quantum of the PXM, and the Threaded Procedure, which groups
closely related Codelets to gain locality during their execution. Within the level of a single system,
the SUs of the Codelet Abstract Machine (CAM) manage creation of TPs and distribution of Codelets.
When looking at larger systems such as an exascale supercomputer, there is a need for a higher level
scheduling mechanism as well. Such work has begun with the creation of DECARD \cite{DEMAC}. 
DECARD fulfills the need of a dedicated scheduling mechanism at the system level that can 
smartly and efficiently distribute TPs throughout a system and manage communications between the
nodes. It is implemented primarily through finite state machines that could be implemented at the
register transfer level and later included as part of the hardware design. DECARD can be viewed 
as an extension of the Codelet Model PXM for better efficiency, load balancing, and performance at
a large scale. The multiple examples mentioned in this section lend credibility to the Codelet
Model PXM on currently used systems, but we now propose that the Codelet Model could provide even
greater benefit on the systems of the future.
\section{The Codelet Model as it Pertains to Chiplets} \label{codelet model for chiplets}
Already, much exposition has been given on the prior work done and the potential benefits provided by a 
fine-grain event-driven PXM like the Codelet Model, even when only implemented as a software runtime. Now, we describe
how the Codelet Model can provide a PXM on chiplet based systems for improved composability, utilization, and memory efficiency.
There are many potential PXMs that could be implemented on a chiplet-based system. Based on prior
trends, it seems probable that a chiplet-based system will rely on an offloading-based PXM, where kernels 
or tasks are pushed to "coprocessors" (which in this case are in fact resident on the host processor chip).
This introduces many problems; though kernels may appear at a glance to be Codelet-like tasks, they are in
reality simply function calls and can have unpredictable behavior like global memory accesses 
or state changes. This model of computation is similar to a tasking PXM, but the underlying
problem is that the PXM implemented at the hardware level has absolutely no concept of a "task", which
allows such a task to have wide communication and side effects. The magnitude of these effects may only
grow in the presence of unified shared memory, because the task can then have direct effects on host memory.
Such unified shared memory may even impose von Neumann abstractions on the memory which would then incur
the same issues that exist on chip multiprocessors now: coherency issues, consistency limitations, unrestricted
behavior, and all the overhead and performance degradation that may accompany them. The Codelet Model can be
a solution in various ways.


\subsection{Programmability and Understandability}
Firstly, chiplets are a powerful example where the Codelet Model can leverage its definition to achieve programmable heterogeneity
with predictable and understandable execution. As long as the behavior meets the standards of the Codelet Model, the 
only necessary difference would be how the Codelets themselves are written by the developer.  However, as the developer
would typically have to explicitly designate code to be executed on the non-CPU cores in something
akin to an offloading or coprocessor model, such code or kernels are
likely well suited to be Codelets (barring the unpredictable behavior mentioned above) 
and the extra development effort would be minimal. 
Then, it follows that the Codelet Model can provide heterogeneous
parallelism, since any Codelets that have their dependencies fulfilled (any enabled Codelet) can be
executed on the respective CU (CPU core, GPU core, etc.). The parallelism inherent in the program
is exposed through the structure of the Codelet Graph and any explicit scheduling or synchronization is no longer the
burden of the developer.
It should be mentioned that in such a case 
as the existing chiplet systems \cite{96Core6Chiplets}\cite{AMDZen2}, the implementation of the Codelet Model on the system will 
be limited (as the systems likely do not have direct hardware support for Codelet Model operations) and 
that due to differences in system layout and memory hierarchy, optimizations should be unique to the platform 
on which they are implemented. Overall, 
the Codelet Model can help maintain programmability at least at a level similar to an offloading-based model. Such is one of the
advantages provided by a clearly defined PXM, but the Codelet Model offers performance benefits as well.

\subsection{Improving Utilization and Memory Efficiency}
The Codelet Model PXM can improve utilization issues, even on a system with various chiplets. Because Codelets have well-defined inputs and outputs,
it is known at compile time what other Codelets will produce the data needed, as well as what other Codelets
will need the data produced. 
The freedom
of being able to organize smaller tasks with clear data dependencies can allow more optimization to occur with the order of memory accesses 
and their collisions (even at the software level \cite{YaoWuThesis}), which will grow more important if
each chiplet or at least various chiplets have their own local memory. The benefits here will be implementation
dependent and rely on the predictability of memory accesses. For example, a scheduling policy could be implemented
that is tailored to avoiding memory bank collisions. Thanks to the graph structure of the program and clearly
defined input and output of Codelets, optimizations could be made which are not feasible on less-defined program
structures like kernels or tasks.

Such clearly defined input/output of Codelets also opens the door to smart scheduling
decisions on the chip. Already, software implementations of the Codelet Model are relatively lightweight and
generally do not incur very much synchronization overhead \cite{DARTSThesis}. Consider that with many distinct
architectures on a chip, it may become difficult to properly distribute tasks in an effective manner. The Codelet Model presents an opportunity
to utilize all chiplets on a chip wherever the parallelism inherent in the problem allows it to, since it indicates
clear dependencies. This allows the scheduling mechanism to make the most of the program graph and allows the developer
to make Codelets as fine-grain as allowed by the algorithm; however, the size of Codelets must be balanced against the 
scheduling/synchronization overhead in terms of utilization. To let Codelets be finer and expose more parallelism in the
system for better utilization, synchronization overhead and scheduling overhead can be reduced
with the help of hardware support.


\subsection{Possible Hardware Augmentations} \label{HW augments}
As mentioned, the Codelet Model PXM can never meet its full potential (nor any other PXM) without hardware
that supports operations specific to the PXM and its corresponding abstract machine. There are many
possible ways to support the Codelet Model in hardware, varying from small differences that have little
effect on conventional parallel computation to those that would have a large
effect on it. There are already some examples of hardware supporting different PXMs that help demonstrate the benefit 
of adding hardware support. As one such example, Tako\cite{tako} includes near-cache reusable dataflow execution engines that
can be programmed at the high-level through definitions of functions such as onMiss (to trigger on
a memory access that registers as a cache miss). What is most interesting here is the clearly exploitable
interface which is provided through overloading functions; in Taco's PXM, operations that occur on cache
misses or cache flushes are part of its abstract machine, though this is not clearly stated, and implemented
in hardware to facilitate high performance. It is an interesting example of how hardware support
for alternative non-conventional PXMs can help improve execution of applications \cite{tako}. It is also
a specific example of the benefits that can only be exploited by tailoring the movement of data throughout
the memory hierarchy to fit an alternative PXM.

As to how the Codelet Model could be augmented through hardware support, there are multiple options. A
simple motivating example is hardware dedicated to tracking Codelet dependencies. In the software based
Codelet Model implementation DARTS, a data structure called a Synchronization Slot is created to hold
a counter of the Codelet's dependencies and decrement it each time another is fulfilled\cite{DARTSThesis}. This could
be implemented in hardware through a counter that has (1) a current dependency count, (2) a reset 
dependency value (which is employed for initialization and Codelet reset for reuse), (3) sufficient
logic to decrement and reset the count, and (4) the ability to receive decrement and reset signals.
 In DARTS, decrementing a Codelet's dependency is a necessarily atomic operation,
meaning that the general flow of the decrement process involves fetching the dependency count value,
decrementing it using the resident ALU, and storing the value back to its location. This process 
takes at a minimum twice the access latency of the dependency count's physical location,
which is ultimately ambiguous both to DARTS and to the user, all while maintaining exclusive access
to the memory location and potentially delaying other signalling further. In the hardware augmented case,
each data structure could receive signals from multiple CUs and serialize them with basic arbitration.
Furthermore, there would be no coherency issues since Codelets are owned by a TP, and TPs belong only
to a single cluster.
This simple mechanism would serve to reduce the signaling overhead when compared to the software 
implementation by reducing the latency of the operation and not requiring exclusive access.

Another possible hardware addition could include hardware Codelet queues. DARTS provides the CUs Codelet
queues from which to pop enabled Codelets to execute, while the SU has a Codelet queue from which to
distribute them\cite{DARTSThesis}. Once more, the physical location of these queues is ambiguous, and with a 
typical multicore cache hierarchy, the queues or their contents may be unexpectedly evicted from a
local cache or invalidated due to operations from other cores (such as pushing a new Codelet to the
queue). These invalidations are costly, and may also result in the data needing to be fetched from
a higher level of the memory hierarchy repeatedly since the queues are accessed often. 
Implementing this queue in hardware should be rather straightforward as Codelets do not
physically contain their code but simply point to it, resulting in all Codelets being of uniform
size and structure. This idea could further be expanded to include a prefetching mechanism for the code
itself, though this may be implementation dependent. As the scheduling policy may also be implementation 
dependent or even reconfigurable, in that case
the queue should be sufficiently flexible to provide the operations needed at the hardware level;
the queue should be well implemented to support the scheduling policies that are available in the
given platform and implementation.

Lastly, as we have already seen, Codelet Model implementations have benefited greatly from including
data streaming mechanisms to support pipelining of Codelet execution\cite{SidThesis}. This could be implemented using
FIFO (First In First Out) queues that support general push and pop interfaces. When used similarly to
\cite{SidThesis}, access to the queues could bypass cache mechanisms entirely and relieve the burden
of the cache coherency mechanism during simultaneous fine-grain accesses to closely-located memory
locations. If the push and pop interfaces are blocking, then there is also the opportunity to avoid
polling mechanisms which are often present in blocking push and pop calls to software implemented
FIFO queues. This can help increase efficiency and speed while still providing fine-grain synchronization
between two or more concurrent Codelets that would otherwise be forced to execute sequentially.

\subsection{The Case Against Cache Coherency}
As mentioned previously, much of the prior research done on chiplet-based chips is related
to interconnect technology and protocols \cite{ModularRoutingDesign}\cite{KiteChipletInterposer}. As a consequence, the discussion
on cache coherency must be mentioned, since cache coherency mechanisms necessitate low latency
interconnects to reduce their effect on shared memory multiprocessing 
and this naturally affects the design of interconnects for chiplets \cite{KiteChipletInterposer}.
We propose removing coherency at specific or all levels of the memory hierarchy 
to lessen the traffic on the interconnects and allow them to dedicate more resources to
data transfer instead of inter-cache coherency messaging. While this certainly will seem radical, it
is significantly more tractable under the implementation of the Codelet Model with hardware support. 
In support of this idea, note that since multiple TPs
do not share the same data elements concurrently, coherency mechanisms across TPs are not needed at all and
could be completely removed if TPs had their own level in the memory hierarchy. The only data
that is conceptually shared between TPs is the results of one TP that another TP requires as
input; this data transfer could be achieved through a message-passing manner that is implicit as part of the
continuation codelet instead of 
utilizing typical shared memory mechanisms, avoiding the coherency problem altogether. This data
transfer between TPs could also safely be achieved through I-Structures\cite{CullerDataflow1986}.

At the Codelet level, no two Codelets that write or overwrite the same data element can 
execute at the same time by definition; in this way, concurrently executing Codelets are
effectively isolated in terms of data accesses (except in the case of pipelining, though
this problem is then resolved through hardware FIFO queues). Generally speaking, Codelets
should then require no coherency mechanism in any memory structure that is not shared
(such as a conventional L1 data cache). In such a case, new write-back policies could be
developed based on Codelet Model semantics between Codelet-level memory and TP-level memory.
Though this is a much more significant
divergence from conventional PXMs in the sense that it has the potential to break most
non-Codelet Model programs run on the system, it would be palatable to add configuration
of the cache such that it can toggle active coherency mechanisms. Another alternative is that
caches on chip could have the possibility to be configured as scratchpad memory much like in
the Polarfire SoC chip. \cite{PolarfireSoCTechRef}. This would allow the Codelet Model PXM to 
leverage it as a fast, local memory that could be used intelligently based on Codelet Model 
semantics for data reuse, prefetching, and movement.  If caches
could be reimagined as portions of scratchpad memory that are configured as having
frame allocation for each Codelet, mechanisms could arise that facilitate more direct
transfer of input and output between Codelets. 
Using hardware-implemented lightweight queue structures could bypass the cache mechanism
and move much of the data transfer or data "writing" to actually passing the new data
directly to the Codelet. 
In other words, shared caches
could be replaced by more manipulable counterparts in order to leverage the semantics
of the Codelet Model for more efficient memory management at the level of a TP.

Beyond mechanisms that particularly rely on the semantics of the Codelet Model, other
ideas can be considered such as the Location Consistency model \cite{LocationConsistency} \cite{EnhancedLocationConsistency}.
The (Enhanced) Location Consistency model has been shown to be weaker than currently used models 
while still producing results equivalent to sequential consistency in data race free programs (which 
well-formed Codelet Model programs are).

Specific implementations, protocols, or scheduling policies in
support of this idea are out of the scope of the paper and should be extrapolated on in
other publications. However, we believe this is an interesting way to improve performance especially considering
the memory wall while maintaining composability and programmability.

\section{Codelet Model on Chiplets Case Study} \label{case study}
In order to give tangible support for the ideas expressed in this paper, a small
case study has been developed to demonstrate the use of the Codelet Model in 
modern problems. A small Python-based Codelet Model simulator that assumes 
ideal scheduling and abundant bandwidth allows us to explore how the Codelet
Model behaves with the inclusion of pipelining and chiplets. While the simulator
is high-level and makes some assumptions, it does follow the execution of a an
actual Codelet Graph that could be used in existing Codelet Model implementations.
After this is further discussed, explanation will be given with respect to the 
limitations of the simulator.
\subsection{Codelet Model Simulator}
The simulator used for this case study can be seen as a sequential, centralized
scheduler of Codelets that replaces execution time of Codelets with a defined delay. It
keeps track of each Codelet's dependency counter and modifies them when prior Codelets
are completed. The signalling between Codelets incurs no delay and because the scheduling
mechanism is ideal, the memory needed for the Codelet is assumed to be already resident
to the CU it is executing on. The scheduling mechanism is also assumed to not incur
overhead. 

The simulator can be run with various options, including the number of CUs and whether
or not it is utilizing pipelining/streaming and chiplets. When executing the simulator using
chiplets enabled, the number of non-chiplet CUs should be adjusted so that it does not
in effect "widen" the hardware that is being simulated. The simulator aims to show the
advantage of using chiplets on the Codelet Model, and this would be skewed by the 
addition of compute resources instead of the replacement of compute resources with
ones tailored to specific applications. 

Chiplets are implemented as CUs of a specific
type that can only execute Codelets enabled for that chiplet type. A "special" Codelet
targeting a specific chiplet applies a related speedup multiplier, reducing the time
spent active before signalling the next Codelets in the graph. In particular, two chiplets
were used in the results shown below: one meant to emulate TPUs\cite{TPU} in their ability 
to perform matrix and vector operations very well, and another meant to emulate the
benefit of accelerators like UDP\cite{UDP2017}, which speeds up branch-heavy tasks such as format
conversion, pattern matching, or compression/decompression. UDP can be embedded as
near-memory compute instead of a traditional accelerator, providing benefits towards
pipelining. The speed up multipliers applied to the Codelets targeting chiplets in the
simulator are 30x and 10x, conservative estimates of TPU\cite{TPUPerformanceAnalysis} and
UDP\cite{UDP2017} speedup respectively to keep them grounded in reality.

Pipelining similar to \cite{SidThesis}, on the other hand, is implemented practically as the ability to concurrently execute 
a Codelet with the Codelet that it depends on.
If a Codelet is marked as pipeline-enabled, upon being enabled it will also enable its
consumer Codelet so that their execution can be overlapped. Pipelining can occur through
an arbitrary number of Codelets.

\subsection{Matrix Multiplication Results}

\begin{figure}
    \centering
    \includegraphics[width=0.4\textwidth]{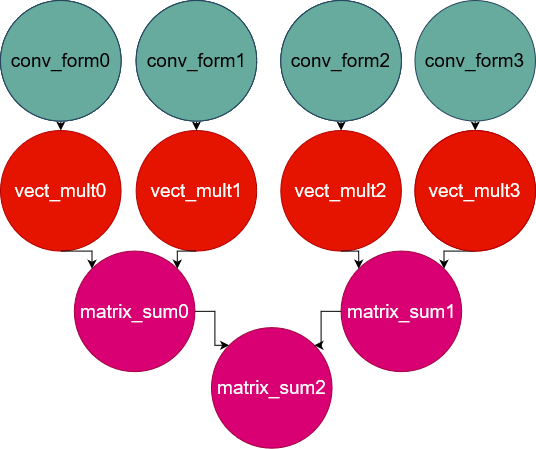}
    \caption[font=scriptsize]{A 4 tile Codelet graph implementing the outer product method of matrix multiplication that is in a sub-optimal format. The Codelet graph converts the matrix to a desirable format, performs multiplication of one column of matrix A and one column of matrix B (at finest grain), producing a partial result matrix which is summed with all other partial result matrices.}
    \label{fig:outer_graph}
\end{figure}

\begin{figure}
\begin{subfigure}{.5\textwidth}
  \centering
  \includegraphics[width=\linewidth]{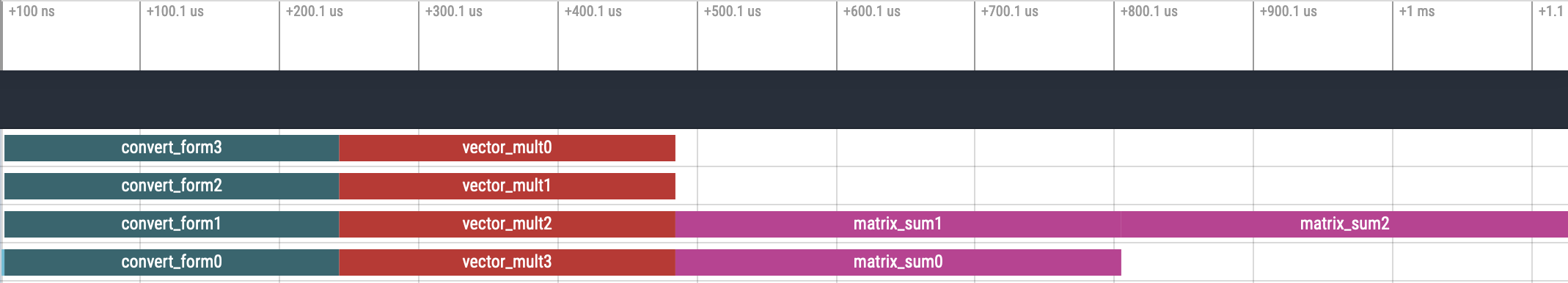}
  \caption{4x4 outer product}
  \label{fig:outer2}
\end{subfigure}
\begin{subfigure}{.5\textwidth}
  \centering
  \includegraphics[width=\linewidth]{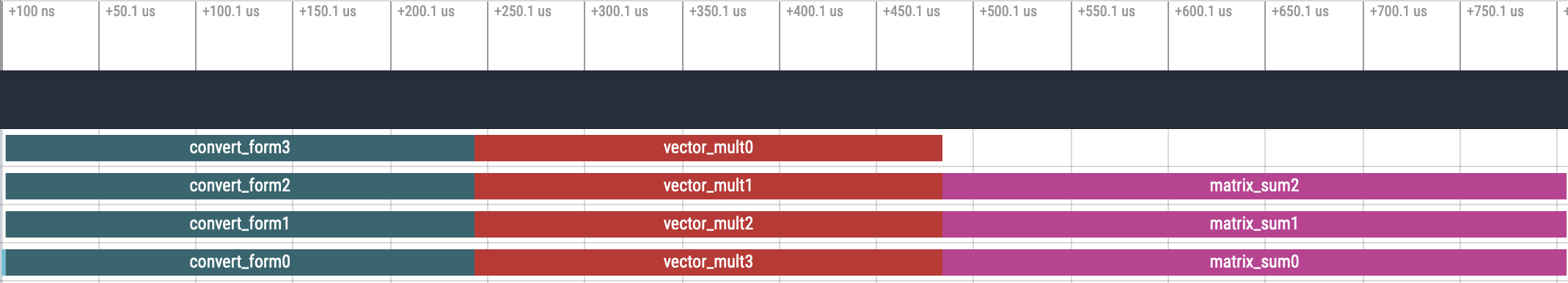}
  \caption{4x4 outer product w/ pipelining}
  \label{fig:outer2_pipe}
\end{subfigure}
\begin{subfigure}{.5\textwidth}
  \centering
  \includegraphics[width=\linewidth]{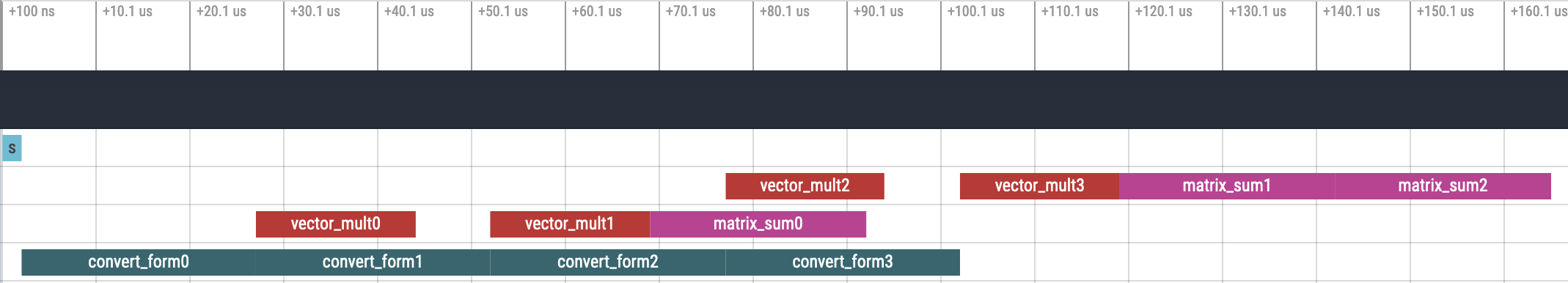}
  \caption{4x4 outer product w/ chiplet accelerators}
  \label{fig:outer2_chip}
\end{subfigure}
\begin{subfigure}{.5\textwidth}
  \centering
  \includegraphics[width=\linewidth]{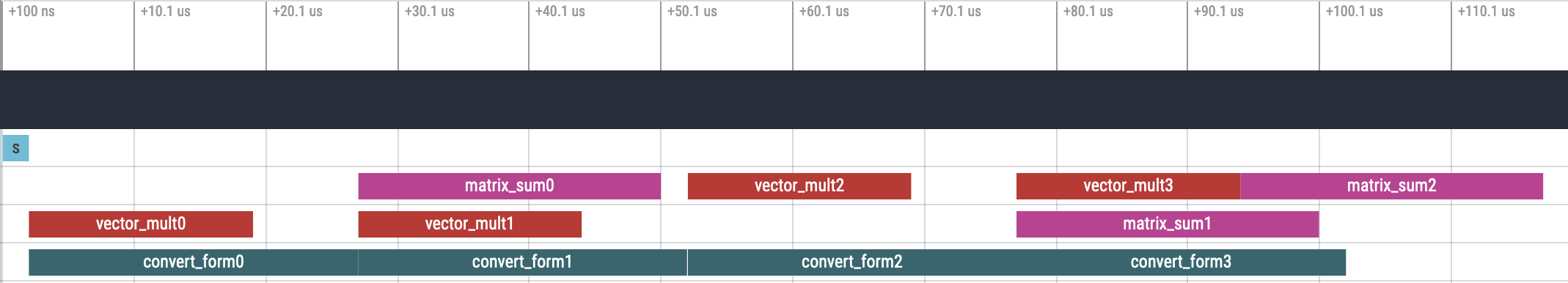}
  \caption{4x4 outer product w/ pipelining and chiplet accelerators}
  \label{fig:outer2_pipe_chip}
\end{subfigure}
\caption{Traces of the outer product method on a 4x4 matrix with 4 CUs. The Codelets executed in these traces are color coordinated with fig. \ref{fig:outer_graph}. These figures are not in the same scale.}
\label{fig:outer_fig}
\end{figure}

Two algorithms for performing GEMM were tested on the simulator: the inner product method
and the outer product method. While less common, the outer product method can present more
opportunities for memory reuse, which can be important when memory access is the limiting factor.
Both methods were tested at various sizes and with varying numbers of compute elements. 
The Codelet graphs for these algorithms are
relatively straight-forward. For example, the inner product method can be realized as a separate
Codelet for each tile in the result matrix, each independent from the other (at smallest tiles possible, each tile represents a single element in the result matrix). One important
note about this is that the program graph for
each GEMM method is effectively based on the number of tiles, and for that reason is not explictly tied to any specific matrix size.
It should be noted that the Codelets would be coarser in a realistic situation (larger tiles) so that the program is not
overly burdened by synchronization and scheduling overheads. The outer product program graph under
test can be seen as a 4 tile example in fig. \ref{fig:outer_graph}. Associated traces of the simulated
execution can be seen in fig. \ref{fig:outer_fig} with various configurations.

The same graph has been expanded up to 64 tiles. Note that as the graph grows wider,
the inverted-tree-like structure of the matrix sum Codelets also grows taller. For example,
the same graph expanded for 8 tiles (Codelets wide) has 7 matrix sum Codelets across 3 "levels"
of the graph. The results of the varying number of tiles can be seen in tables \ref{table:outer_time}
and \ref{table:outer_speedup}. Notice that the version that consistently performs the best is 
with pipelining and chiplet usage combined, but that it begins scaling poorly in the outer product method in terms of speedup
once number of tiles reaches 64. This is because in this example, the two chiplet types are given equal
amounts of resources (about 32 each). Once the tile count reaches 32, the best speedup is shown
because this is where the amount of each resource matches the number of tiles (and therefore, width of the graph) and the pipelining
allows the entire matrix sum portion of the tree to execute concurrently, conceptually streaming
the data throughout the CUs and pipelining the entire sum. After the number of tiles grows to 64, the
speedup is not as substantial because the resources are saturated while they are not in the non-chiplet
cases (since the 64 CUs are all conventional). The speedup can also vary based on the ratio of resources
contained by one chiplet to the other chiplet. Beyond this, part of this change in speedup could be due to the way
the Codelet delays were modeled: the Codelets representing the format conversion and vector multiplication
scale by O(Tiles) while the matrix sum Codelets scale by O(Tiles\textsuperscript{2}), meaning
that the benefit of each chiplet also scales unevenly. 

Alternatively, the speedups achieved for the inner product method are less significant than those achieve for the outer product
method. This is because the inner product method already exposes as much parallelism as possible based on tile width, and as a result
benefits very little from pipelining. Also, it most likely suffers more severely from the limited number of chiplet CUs dedicate to
speeding up the core calculation portion, since this conceptually results in the chiplet version having less CUs than the normal 
version. Having less CUs available to execute the Codelets also further limits the benefit of pipelining in the simulation since
there is no memory behavior being modeled in the simulator (see section \ref{sim justification}).

Furthermore, the sequential centralized simulation grows
increasingly taxing as the tile count increases exponentially. As a result, we are currently limited 
to 64 tiles. Testing different ratios of chiplet resources or by 
making matrix sums finer grain is left for future work.

\begin{table}
\tiny
\centering
\begin{tabular}{||c|c c c c|c c c c||} 
\hline
 & \multicolumn{4}{|c|}{Outer Product (Time Units)} & \multicolumn{4}{|c||}{Inner Product (Time Units)}\\
 \hline
 Tiles & Basic & Pipelined & Chiplets & Both & Basic & Pipelined & Chiplets & Both  \\ [0.5ex] 
 \hline
 8 & 1208 & 324 & 94 & 26 & 204 & 133 & 33 & 22\\ 
 \hline
 16 & 5609 & 1284 & 405 & 93 & 1607 & 984 & 235 & 141\\
 \hline
 32 & 26570 & 5605 & 1930 & 411 & 12819 & 7732 & 1827 & 1094\\
 \hline
 64 & 124811 & 22406 & 9005 & 3376 & 102467 & 61572 & 14467 & 8717\\
 \hline
\end{tabular}
\caption{GEMM results with 64 CUs available. For the chiplet versions, the CUs are divided evenly between the two chiplet types with 1 conventional CU for the start and end Codelets.}
\label{table:outer_time}
\end{table}

\begin{table}
\tiny
\centering
\begin{tabular}{||c|c c c| c c c ||} 
\hline
  & \multicolumn{3}{|c|}{\textbf{Outer Product (Speedup \%)}} & \multicolumn{3}{|c||}{\textbf{Inner Product (Speedup \%)}}  \\
 \hline
 \textbf{Tiles} & \textbf{Pipelined} & \textbf{Chiplets} & \textbf{Both} & \textbf{Pipelined} & \textbf{Chiplets} & \textbf{Both} \\ [0.5ex] 
 \hline
 8 & 3.72 & 12.8 & 46.4 & 1.53 & 6.18 & 9.27 \\ 
 \hline
 16 & 4.36 & 13.8 & 60.3 & 1.63 & 6.84 & 11.4  \\
 \hline
 32 & 4.74 & 13.7 & 64.6 & 1.66 & 7.02 & 11.7  \\
 \hline
 64 & 5.57 & 13.9 & 37.0 & 1.67 & 7.08 & 11.8  \\
 \hline
\end{tabular}
\caption{Speedup(\%) of extensions to the Codelet Model on the GEMM methods. Same configuration as in table \ref{table:outer_time}.}
\label{table:outer_speedup}
\end{table}

\subsection{Simulator Justification and Future Work} \label{sim justification}
It is clear that our current implementation of a Codelet Model simulator for analyzing 
the performance of Codelet programs is limited. However, the results and the simulator
can be justified as it pertains to a full-system Codelet Model
PXM implementation with hardware support. Firstly, critics of the simulator will notice
that it does not account for signalling and scheduling overhead. We posit that the overhead
incurred by signalling can be mitigated through a hardware mechanism like the one mentioned
in section \ref{HW augments}. Latency incurred by that operation in most cases would be 
hidden by the scheduling and execution of other already-enabled Codelets. Similarly, the
scheduling overhead could be mitigated through hardware queues also mentioned in section
\ref{HW augments}. Other scheduling overhead components such as the latency of the
scheduling decisions are left for future work.

One thing lacking in the simulator is memory behavior. With a good enough
scheduling mechanism, we can assume that Codelets are scheduled where their input data
are already resident. Beyond this, we predict that the benefit of pipelining/streaming in this
simulator is underestimated.
In a realistic system, the execution could benefit from HW FIFOs bypassing caching coherency
mechanisms and ensuring low latency access to the data.
In the simulator pipelining effectively only increases the number of 
Codelets that can be enabled at a single moment based on the Codelet Graph which
improves utilization. In the future we expect to improve this
simulator to include memory behavior and use it as a tool to develop better
memory-conscious scheduling decisions. 
\section{Conclusion}
Chiplets are the hardware answer to industry’s constant push for improvement in multiple problem areas, but 
will introduce even more difficulties to software development methods that are already wrought with issues. 
In turn, restrictions and difficulties in software development will limit the efficient use of the hardware 
itself. The answer is a clearly defined PXM integrated at all levels, and the fine-grain event-driven Codelet 
Model PXM presents new opportunities to improve performance and efficiency while maintaining understandability 
and clearly defined interactions. Improvements made to the Codelet Model such as FIFO queues for dataflow 
software pipelining can be implemented at the hardware level, helping to improve utilization and limit memory 
access latency within Codelets. In the future, we hope to further show the 
validity of the Codelet Model as a solution through improving the simulator to include memory behavior and a 
wider array of applications. While utilization and speedup speak for themselves through simulation, other 
stated benefits like understandability and composability of programs cannot be quantified, but will be known 
to all those who have spent sleepless nights debugging a multithreaded program.

\bibliographystyle{IEEEtranS}
\bibliography{bib/capsl, bib/CodeletModelforChiplets}

\end{document}